\documentclass[a4paper,12pt]{article}
\usepackage{amsfonts}
\usepackage{lscape}
\begin{document}

\makeatletter
\@addtoreset{equation}{section}
\makeatother
\renewcommand{\theequation}{\thesection.\arabic{equation}}

\def\tr#1{\mbox{tr}\,#1}

\def\cosx#1{{\tt c}{\left(#1\right)}}
\def\coshx#1{{\tt K}{\left(#1\right)}}
\def\sinhx#1{{\tt S}{\left(#1\right)}}
\def\psip#1{\psi^+_{#1}}
\def\psim#1{\psi^-_{#1}}
\def\psipm#1{\psi^\pm_{#1}}

\thispagestyle{empty}
\begin{flushright}
AEI-2002-004
\end{flushright}

\begin{center}
{\bf\Large Some stationary points \\ of gauged N=16 D=3 supergravity}\\
\bigskip\bigskip\bigskip\bigskip
{\bf T. Fischbacher\medskip\\ }
{\em Max-Planck-Institut f\"ur Gravitationsphysik,\\
     Albert-Einstein-Institut,\\
     M\"uhlenberg 1, D-14476 Potsdam, Germany\\ }
\smallskip
{\small \tt tf@aei-potsdam.mpg.de}\bigskip

\end{center}

\begin{abstract}
\noindent Five nontrivial stationary points are found for maximal gauged N=16 supergravity in
three dimensions with gauge group $SO(8)\times SO(8)$ by restricting
the potential to a submanifold of the space of 
$SU(3)\subset\left(SO(8)\times SO(8)\right)_{\rm diag}$ singlets.
The construction
presented here uses the embedding of $E_{7(+7)}\subset E_{8(+8)}$ to
lift the analysis of $N=8, D=4$ supergravity performed by N. Warner to
$N=16, D=3$, and hence, these stationary points correspond to some of
the known extrema of gauged $N=8, D=4$ supergravity.
\end{abstract}

\vfill
\leftline{{January 2002}}

\newpage

\section{Introduction}

In $N>1$ extended supergravity theories in dimensions $D\ge4$, it is
possible to gauge the $SO(N)$ symmetry that rotates the supersymmetry
generators into each other. In three dimensions, the case of gauged
$N=16$ supergravity is of particular interest, since it exhibits a
very rich structure; here, scalar fields are on-shell equivalent to
vector fields, and due to the freedom in the choice of the number of
vector fields defined as nonlocal functions of the scalar fields, a
large number of gauge groups become possible
\cite{Nicolai:2000sc, Nicolai:2001sv}. Furthermore, in contrast to
$D\ge4$ supergravity, maximal three-dimensional gauged supergravities
are not derivable by any technique known so far from any of the known
higher-dimensional maximal gauged supergravity theories, since the
vector fields show up in the Lagrangian via a Chern-Simons term and
not via a kinetic term as in a Lagrangian obtained by Kaluza-Klein
compactification.

Gauging of any extended supergravity introduces into the Lagrangian,
among other terms, at second order in the gauge coupling constant~$g$
a potential for the scalar fields. For $D=4$, $N=2,3$, the scalar
potential is just a cosmological constant $-6\,g^2$, while for $N\ge
5$, it features a rich extremal structure which defies
an exhaustive analysis for $N=8$. Some nontrivial extrema of $D=4, N=8$
supergravity have been determined in
\cite{Warner:du, Warner:vz} by employing group-theoretical arguments.
In order to shed light on the question how the extremal structure of
maximal gauged $D=3, N=16$ supergravity constructed in
\cite{Nicolai:2000sc, Nicolai:2001sv} is related to that of $D=4, N=8$
when one chooses $SO(8)\times SO(8)$ as gauge group, it is interesting
to try to generalize the construction given in \cite{Warner:vz} to
this case.

The potential of gauged maximal $N=16$ supergravity with maximal
compact $SO(8)\times SO(8)$ gauge group which we investigate here is
considerably more complicated than any supergravity potential
previously considered and may well be the most complicated analytic
potential ever studied. The significance of the deeply involved
structure of the exceptional Lie group $E_8$ showing up in this case
still remains to be elucidated.

\section{The scalar potential}

The space of the $128$ scalars of $N=16, D=3$ supergravity can be
identified with the symmetric space $E_{8(+8)}/SO(16)$, with $SO(16)$
being the maximal compact subgroup of $E_{8(+8)}$ and
as $E_{8(+8)}$ is obtained by fusing the adjoint representation of
$SO(16)$ with the Majorana-Weyl spinor representation of $SO(16)$, we
split $E_8$ indices $\mathcal{A}, \mathcal{B},\ldots$ via
$\mathcal{A}=(A,[IJ])$, where indices $A,B,\ldots$ denote $SO(16)$
spinors and indices $I,J,\ldots$ belong to the fundamental
representation of $SO(16)$. The structure constants of $E_{8(+8)}$ are
\begin{equation}
f_{IJ\;KL}{}^{MN}=-8\,\delta_{{}_[I{}^[K}^{\phantom{MN}}\delta_{L{}^]J{}_]}^{MN},\qquad
f_{IJ\,A}{}^{B}={\textstyle \frac{1}{2}}\,\Gamma^{IJ}_{AB}.
\end{equation}

The rationale of the common convention to introduce an extra factor
$1/2$ for every antisymmetric index pair $[IJ]$ that is summed over is
explained in the appendix.

From the $E_{8(+8)}$ matrix generators
$t_\mathcal{A}{}^{\mathcal{C}}{}_{\mathcal{B}}=f_{\mathcal{A}\mathcal{B}}{}^{\mathcal{C}}$,
one forms the Cartan-Killing metric
\begin{equation}
\eta_{\mathcal{A}\mathcal{B}}={\textstyle \frac{1}{60}}\tr t_\mathcal{A}t_\mathcal{B};\qquad\eta_{AB}=\delta_{AB},\quad\eta_{IJ\;KL}=-2\delta^{IJ}_{KL}.
\end{equation}

In order to obtain the potential, we first introduce the
zwei\-hundert\-achtund\-vierzig\-bein $\mathcal{V}$ in an
unitary gauge via
\begin{equation}
\mathcal{V}=\exp\,\left(\psi^A t_A\right)
\end{equation}
where $\psi^A$ is a $SO(16)$ Majorana-Weyl-Spinor and
$t_A$ are the corresponding generators of $E_{8(+8)}$.

Due to the on-shell equivalence of scalars and vectors in three
dimensions, the structure of gauged maximal $D=3, N=16$ supergravity
is much richer than in higher dimensions; here, besides the maximal
compact gauge group $SO(8)\times SO(8)$ and its noncompact forms, it
is possible to also have a variety of noncompact exceptional gauge groups.
The choice of gauge group $G_0$ is parametrized by the Cartan-Killing
metric of $G_0$ embedded in $E_{8(+8)}$. The requirement of maximal
supersymmetry reduces to a single algebraic condition for this
symmetric tensor $\Theta$ which states that it must not have a
component in the ${\bf 27000}$ of the $E_{8(+8)}$ tensor product
decomposition $\left({\bf 248}\times {\bf 248}\right)_{\rm sym}={\bf 1}+{\bf 3875}+{\bf 27000}$.
Obviously, one extremal case is $G_0=E_{8(+8)}$. In this case, the scalar potential
again reduces to just a cosmological constant, but the smaller we
choose the gauge group, the richer the extremal structure of the
corresponding potential becomes.

From this embedding tensor $\Theta_{\mathcal MN}$, the $T$-tensor now
is formed by
\begin{equation}
T_{\mathcal{AB}}={\mathcal{V}}{}^{\mathcal M}{}_{\mathcal A}{\mathcal{V}}{}^{\mathcal N}{}_{\mathcal B}\Theta_{\mathcal{MN}}.
\end{equation}

With $\theta=\frac{1}{248}\eta^{\mathcal KL}\Theta_{\mathcal KL}$, we
form the tensors
\begin{equation}
\begin{array}{lll}
A_1^{IJ}&=&\frac{8}{7}\,\theta\delta_{IJ}+\frac{1}{7}T_{IK\,JK}\\
A_2^{I\dot A}&=&-\frac{1}{7}\Gamma^{J}_{A\dot A}T_{IJ\,A}\\
A_3^{\dot A\dot B}&=&2\theta\delta_{\dot A\dot B}+\frac{1}{48}\Gamma^{IJKL}_{\dot A\dot B}T_{IJ\,KL}.
\end{array}
\end{equation}

The potential is given by
\begin{equation}
V\left(\psi^A\right)=-\frac{1}{8}\,g^2\left(A_1^{IJ}A_1^{IJ}-\frac{1}{2}A_2^{I\dot A}A_2^{I\dot A}\right).
\end{equation}

If we split the $SO(16)$ vector index $I$ into $SO(8)$ Indices
$I=(i,\bar j)$, the nonzero components of the embedding tensor for
the gauge group $SO(8)\times SO(8)$ considered here are given by
\begin{equation}
\Theta_{ij\,kl}=8\,\delta^{ij}_{kl},\qquad \Theta_{\overline{ij}\,\overline{kl}}=-8\,\delta^{\overline{ij}}_{\overline{kl}}
\end{equation}
where we use the same normalization as in \cite{Nicolai:2001sv}.

The most fruitful technique for a study of the extremal structure of
these potentials known so far appears to be that introduced in
\cite{Warner:vz}: first, choose a subgroup $H$ of the gauge group~$G$
($SO(8)$ for $N=8, D=4$, 
$SO(8)\times SO(8)$ in for the case considered here); then, determine
a parametrization of the submanifold $M$ of $H$-singlets of the
manifold of physical scalars $P$. Every point of this submanifold for
which all derivatives within $M$ vanish must also have vanishing
derivatives within $P$. The reason is that, with the potential $V$
being invariant under $G$ and hence also under $H$, the power series
expansion of a variation $\delta z$ of $V$ around a stationary point
$z_0$ in $M$ where $\delta z$ points out of the submanifold $M$ of
$H$-singlets can not have a $\mathcal{O}(\delta z)$ term, since each
term of this expansion must be invariant under $H$ and it is not
possible to form a $H$-singlet from just one $H$-nonsinglet. All the
stationary points found that way will break the gauge group down to a
symmetry group that contains $H$.

The general tendency is that, with $H$ getting smaller, the number of
$H$-singlets among the supergravity scalars will increase. For
$H$-singlet spaces of low dimension, it easily happens that the scalar
potential does not feature any nontrivial stationary points at all,
while for higher-dimensional singlet spaces, the potential soon
becomes intractably complicated. Using the embedding of $SU(3)\subset
SO(8)$ under which the scalars, vectors and co-vectors of $SO(8)$
decompose into ${\bf 3}+\bar {\bf 3}+{\bf 1}+{\bf 1}$ for $N=8, D=4$
gives a case of manageable complexity with six-dimensional scalar
manifold for which five nontrivial extrema were given in a complete
analysis in
\cite{Warner:vz}. (It seems reasonable to expect further
yet undiscovered extrema breaking $SO(8)$ down to groups smaller than
$SU(3)$.)

Since it is interesting to see how the extremal structure of $N=8,
D=4$ gauged $SO(8)$ supergravity is related to $N=16, D=3$
$SO(8)\times SO(8)$ gauged supergravity, it is reasonable to try to
lift the construction given in \cite{Warner:vz} to this case via the
embedding of $E_{7(+7)}$ in $E_{8(+8)}$ described in the appendix. As
explained there in detail, the $128$ spinor components $\psi^A$
decompose into $2\times 1$ $SO(8)$ scalars which we call $\psipm{}$,
$2\times 28$ 2-forms $\psipm{{\tt i}_1 {\tt i}_2}$, and
$2\times 35$ 4-forms $\psipm{{\tt i}_1 {\tt i}_2 {\tt i}_3{\tt i}_4}$.%
\footnote{Here, ${\tt i}, {\tt k},\ldots$ denote
$SO(8)$ indices of the diagonal $SO(8)$ of $SO(8)\times SO(8)$.}

Re-identifying the $E_{8(+8)}$ generators corresponding to the $SU(3)$
singlets, resp. the $SU(8)$ rotations used to parametrize the singlet
manifold given in \cite{Warner:vz} is straightforward; exponentiating
them, however, is not. Looking closely at explicit $248\times248$
matrix representations of these generators reveals that, after
suitable re-ordering of coordinates, they decompose into blocks of
maximal size $8\times8$ and are (by using a computer) sufficiently
easy to diagonalize.  Considerable simplification of the task of
computing explicit analytic expressions for the scalar potential by
making use of as much group theoretical structure as possible is
expected, but nowadays computers are powerful enough to allow a
head-on approach using explicit 248-dimensional component notation and
symbolic algebra on sparsely occupied tensors.\footnote{Maple as well
as Mathematica do not perform well enough here, while FORM only has
comparatively poor support for tensors in explicit component
representation; hence, all the symbolic algebra was implemented from
scratch using the CMUCL Common LISP compiler.} The original motivation
to invest time into the design of aggressively optimized explicit
symbolic tensor algebra code comes from the wealth of different cases
due to the large number of possible gauge groups of $D=3, N=16$
supergravity.

One important complication arises from the fact that the $({\bf 56},{\bf 2})$ and
$({\bf 1},{\bf 3})$ representations give additional $SU(3)$ singlets: from the
$({\bf 1},{\bf 3})$, these are the generators corresponding to $\psipm{}$, while
each of the 2-forms $F^\pm$ which are defined as intermediate
quantities in \cite{Warner:vz} appears twice (once for $\psip{{\tt
i}_1{\tt i}_2}$, once for $\psim{{\tt i}_1{\tt i}_2}$), giving a total
of six extra scalars, so our scalar manifold $M$ now is
$12$-dimensional. While explicit analytic calculation of the potential
on a submanifold of $M$ reveals that the full 12-dimensional potential
definitely is way out of reach of a complete analysis using standard
techniques, it is nevertheless possible to make progress by making
educated guesses at the possible locations of extrema; for example,
one notes that for four of the five stationary points given in
\cite{Warner:vz}, the angular parameters are just such that the sines
and cosines appearing in the potential all assume values
$\{-1;0;+1\}$. Hence it seems reasonable to try to search for
stationary points by letting these compact coordinates run through a
discrete set of special values only, thereby reducing the number of
coordinates.

The immediate problem with the consideration of only submanifolds $M'$
of the full manifold $M$ of singlets is that, aside from not being
able to prove the nonexistence of further stationary points on $M$,
the vanishing of derivatives within $M'$ does not guarantee to have a
stationary point of the full potential. A sieve for true solutions is
given by the stationarity condition (4.12) in \cite{Nicolai:2001sv}:
\begin{equation}
\label{SolutionSieve}
3\,A_1^{IM}A_2^{M\dot A}=A_2^{I\dot B}A_3^{\dot A\dot B}.
\end{equation}

\vfill

\section{The SU(3) singlets}

Explicitly, the 12 singlets are
\begin{itemize}

\item The 6 selfdual complex 4-form singlets in the $({\bf 133},{\bf 1})$ (using the same terminology as in \cite{Warner:vz}):
\begin{equation}
\begin{array}{lll}
G_1^+&=&\left(\psim{1234}+\psim{1256}+\psim{1278}\right)^A\,t_A\\
G_1^-&=&\left(\psip{1234}+\psip{1256}-\psip{1278}\right)^A\,t_A\\
G_2^+&=&\left(-\psim{1357}+\psim{1368}+\psim{1458}+\psim{1467}\right)^A\,t_A\\
G_2^-&=&\left(\psip{1357}-\psip{1467}+\psip{1458}+\psip{1368}\right)^A\,t_A\\
G_3^+&=&\left(-\psim{1468}+\psim{1367}+\psim{1358}+\psim{1457}\right)^A\,t_A\\
G_3^-&=&\left(\psip{1468}-\psip{1358}+\psip{1367}+\psip{1457}\right)^A\,t_A
\end{array}
\end{equation}

\item The two scalars from $({\bf 1},{\bf 3})$:
\begin{equation}
\begin{array}{lll}
S_1&=&\psi^{-A}_{}\,t_A\\
S_2&=&\psi^{+A}_{}\,t_A
\end{array}
\end{equation}

\item The spinors corresponding to the two-forms $F^\pm$ of \cite{Warner:vz},
once built from the $\psip{{\tt i}_1{\tt i}_2}$,
and once built from the $\psim{{\tt i}_1{\tt i}_2}$:
\begin{equation}
\begin{array}{lll}
F^{++}&=&\left(\psip{12}+\psip{34}+\psip{56}+\psip{78}\right)^A\,t_A\\
F^{+-}&=&\left(\psip{12}+\psip{34}+\psip{56}-\psip{78}\right)^A\,t_A\\
F^{-+}&=&\left(\psim{12}+\psim{34}+\psim{56}+\psim{78}\right)^A\,t_A\\
F^{--}&=&\left(\psim{12}+\psim{34}+\psim{56}-\psim{78}\right)^A\,t_A
\end{array}
\end{equation}
\end{itemize}

Using the same parametrization of the six singlets from $({\bf 133},{\bf 1})$ as
in \cite{Warner:vz}, that is, writing them as

\begin{equation}
\label{ManifoldSpinorsSix}
\psi=S\left(\lambda_1\,G_1^+ + \lambda_2\,G_2^+\right)
\end{equation}

\vfill\break

with
\begin{equation}
\begin{array}{lll}
S&=&\mbox{diag}\left(\omega,\omega,\omega,\omega,\omega,\omega,\omega^{-3}P\right),\qquad \omega=e^{ia/4},\\
&&\\
P&=&
\left(\begin{array}{rr}\cos\vartheta&-\sin\vartheta\\ \sin\vartheta&\cos\vartheta\end{array}\right)
\left(\begin{array}{rr}e^{i\varphi}&0\\0&e^{-i\varphi}\end{array}\right)
\left(\begin{array}{rr}\cos\Psi&-\sin\Psi\\\sin\Psi&\cos\Psi\end{array}\right),
\end{array}
\end{equation}
and using the translation of $SU(8)$ generators to $E_{8(+8)}$
generators of the appendix, the formula corresponding to $(2.10)$ in
\cite{Warner:vz} reads%
\begin{equation}
\begin{array}{lll}
-8\,g^{-2}V&=&\frac{189}{2}-
\frac{3}{2}\,\coshx{3 \lambda_1} \coshx{4 \lambda_2} \cosx{2 a} \cosx{2 \varphi}\\
&&+\frac{3}{2}\,\coshx{3 \lambda_1} \cosx{2 a} \cosx{2 \varphi}
+12\,\coshx{2 \lambda_1} \coshx{2 \lambda_2} \cosx{2 a} \cosx{2 \varphi}\\
&&-12\,\coshx{2 \lambda_1} \cosx{2 a} \cosx{2 \varphi}
+\frac{3}{2}\,\coshx{ \lambda_1} \coshx{4 \lambda_2} \cosx{2 a} \cosx{2 \varphi}\\
&&-\frac{3}{2}\,\coshx{ \lambda_1} \cosx{2 a} \cosx{2 \varphi}
-12\,\coshx{2 \lambda_2} \cosx{2 a} \cosx{2 \varphi}\\
&&+12\,\cosx{2 a} \cosx{2 \varphi}
+\frac{23}{8}\,\coshx{3 \lambda_1}
-\frac{19}{8}\,\coshx{3 \lambda_1} \coshx{4 \lambda_2}\\
&&-\frac{1}{8}\,\coshx{3 \lambda_1} \coshx{4 \lambda_2} \cosx{4 \varphi}
-\frac{1}{2}\,\coshx{3 \lambda_1} \coshx{2 \lambda_2}\\
&&+\frac{1}{2}\,\coshx{3 \lambda_1} \coshx{2 \lambda_2} \cosx{4 \varphi}
-\frac{3}{8}\,\coshx{3 \lambda_1} \cosx{4 \varphi}\\
&&+12\,\coshx{2 \lambda_1}
+36\,\coshx{2 \lambda_1} \coshx{2 \lambda_2}
+\frac{405}{8}\,\coshx{ \lambda_1}\\
&&-\frac{9}{8}\,\coshx{ \lambda_1} \coshx{4 \lambda_2}
-\frac{3}{8}\,\coshx{ \lambda_1} \coshx{4 \lambda_2} \cosx{4 \varphi}\\
&&+\frac{93}{2}\,\coshx{ \lambda_1} \coshx{2 \lambda_2}
+\frac{3}{2}\,\coshx{ \lambda_1} \coshx{2 \lambda_2} \cosx{4 \varphi}\\
&&-\frac{9}{8}\,\coshx{ \lambda_1} \cosx{4 \varphi}
+\frac{7}{2}\,\coshx{4 \lambda_2}
+\frac{1}{2}\,\coshx{4 \lambda_2} \cosx{4 \varphi}\\
&&+14\,\coshx{2 \lambda_2}
-2\,\coshx{2 \lambda_2} \cosx{4 \varphi}
+\frac{3}{2}\,\cosx{4 \varphi}.
\end{array}
\end{equation}

Here and in what follows, we use the abbreviations
\begin{equation}
\cosx{\alpha}=\cos(\alpha),\qquad \coshx{\sigma}=\cosh(\sigma),\qquad \sinhx{\tau}=\sinh(\tau).
\end{equation}

Note that just as in \cite{Warner:vz}, we have
$\vartheta$-independence due to $SO(8)$ invariance {\em as well as} an
additional independence of $\Psi$.

A detailed analysis of this restricted potential shows that there
are seven candidates for nontrivial stationary points, but none of these
is a true stationary point of the full potential. However, it is
observed numerically that the derivative at many of these points lies
in the $\psipm{}$ plane, hence we extend
(\ref{ManifoldSpinorsSix}) to
\begin{equation}
\psi=S\left(\lambda_1\,G_1^+ + \lambda_2\,G_2^+\right)+\sigma_1\,S_1
\end{equation}
and obtain the potential
\begin{equation}
\begin{array}{lll}
-8g^{-2}V&=&
\frac{189}{2}
+\frac{1}{8}\,\sinhx{3 \lambda_1} \coshx{4 \lambda_2} \sinhx{ \sigma_1} \cosx{3 a} \cosx{4 \varphi}\\
&&+\frac{1}{8}\,\sinhx{3 \lambda_1} \coshx{4 \lambda_2} \sinhx{ \sigma_1} \cosx{3 a}
-\frac{1}{8}\,\coshx{3 \lambda_1} \coshx{4 \lambda_2} \coshx{ \sigma_1} \cosx{4 \varphi}\\
&&-\frac{1}{2}\,\sinhx{3 \lambda_1} \coshx{2 \lambda_2} \sinhx{ \sigma_1} \cosx{3 a} \cosx{4 \varphi}
+\frac{1}{2}\,\sinhx{3 \lambda_1} \coshx{2 \lambda_2} \sinhx{ \sigma_1} \cosx{3 a}\\
&&+\frac{3}{8}\,\sinhx{3 \lambda_1} \sinhx{ \sigma_1} \cosx{3 a} \cosx{4 \varphi}
-\frac{5}{8}\,\sinhx{3 \lambda_1} \sinhx{ \sigma_1} \cosx{3 a}\\
&&-\frac{3}{8}\,\sinhx{ \lambda_1} \coshx{4 \lambda_2} \sinhx{ \sigma_1} \cosx{3 a} \cosx{4 \varphi}
-\frac{3}{8}\,\sinhx{ \lambda_1} \coshx{4 \lambda_2} \sinhx{ \sigma_1} \cosx{3 a}\\
&&-\frac{3}{8}\,\coshx{ \lambda_1} \coshx{4 \lambda_2} \coshx{ \sigma_1} \cosx{4 \varphi}
+\frac{3}{2}\,\sinhx{ \lambda_1} \coshx{2 \lambda_2} \sinhx{ \sigma_1} \cosx{3 a} \cosx{4 \varphi}\\
&&-\frac{3}{2}\,\coshx{3 \lambda_1} \coshx{4 \lambda_2} \coshx{ \sigma_1} \cosx{2 a} \cosx{2 \varphi}\\
&&+\frac{3}{2}\,\coshx{ \lambda_1} \coshx{4 \lambda_2} \coshx{ \sigma_1} \cosx{2 a} \cosx{2 \varphi}\\
&&+\frac{3}{2}\,\sinhx{3 \lambda_1} \coshx{4 \lambda_2} \sinhx{ \sigma_1} \cosx{a} \cosx{2 \varphi}\\
&&+\frac{3}{2}\,\sinhx{ \lambda_1} \coshx{4 \lambda_2} \sinhx{ \sigma_1} \cosx{a} \cosx{2 \varphi}\\
&&-\frac{3}{4}\,\sinhx{ \lambda_1} \coshx{4 \lambda_2} \sinhx{ \sigma_1} \cosx{a}
-\frac{3}{8}\,\coshx{3 \lambda_1} \coshx{ \sigma_1} \cosx{4 \varphi}\\
&&-\frac{3}{2}\,\sinhx{ \lambda_1} \coshx{2 \lambda_2} \sinhx{ \sigma_1} \cosx{3 a}
+\frac{3}{2}\,\coshx{3 \lambda_1} \coshx{ \sigma_1} \cosx{2 a} \cosx{2 \varphi}\\
&&-\frac{3}{2}\,\coshx{ \lambda_1} \coshx{ \sigma_1} \cosx{2 a} \cosx{2 \varphi}
-\frac{3}{2}\,\sinhx{3 \lambda_1} \sinhx{ \sigma_1} \cosx{a} \cosx{2 \varphi}\\
&&+\frac{3}{2}\,\coshx{ \lambda_1} \coshx{2 \lambda_2} \coshx{ \sigma_1} \cosx{4 \varphi}
-\frac{9}{8}\,\sinhx{ \lambda_1} \sinhx{ \sigma_1} \cosx{3 a} \cosx{4 \varphi}\\
&&+\frac{15}{8}\,\sinhx{ \lambda_1} \sinhx{ \sigma_1} \cosx{3 a}
+12\,\coshx{2 \lambda_1} \coshx{2 \lambda_2} \cosx{2 a} \cosx{2 \varphi}\\
&&-24\,\sinhx{ \lambda_1} \coshx{2 \lambda_2} \sinhx{ \sigma_1} \cosx{a} \cosx{2 \varphi}
+\frac{3}{2}\,\cosx{4 \varphi}\\
&&-12\,\coshx{2 \lambda_1} \cosx{2 a} \cosx{2 \varphi}
-12\,\coshx{2 \lambda_2} \cosx{2 a} \cosx{2 \varphi}\\
&&-24\,\sinhx{ \lambda_1} \coshx{2 \lambda_2} \sinhx{ \sigma_1} \cosx{a}
+12\,\cosx{2 a} \cosx{2 \varphi}\\
&&+\frac{9}{4}\,\sinhx{3 \lambda_1} \coshx{4 \lambda_2} \sinhx{ \sigma_1} \cosx{a}
-\frac{9}{4}\,\sinhx{3 \lambda_1} \sinhx{ \sigma_1} \cosx{a}\\
&&+\frac{45}{2}\,\sinhx{ \lambda_1} \sinhx{ \sigma_1} \cosx{a} \cosx{2 \varphi}
+\frac{99}{4}\,\sinhx{ \lambda_1} \sinhx{ \sigma_1} \cosx{a}\\
&&-\frac{19}{8}\,\coshx{3 \lambda_1} \coshx{4 \lambda_2} \coshx{ \sigma_1}
+\frac{1}{2}\,\coshx{3 \lambda_1} \coshx{2 \lambda_2} \coshx{ \sigma_1} \cosx{4 \varphi}\\
&&-\frac{1}{2}\,\coshx{3 \lambda_1} \coshx{2 \lambda_2} \coshx{ \sigma_1}
+\frac{23}{8}\,\coshx{3 \lambda_1} \coshx{ \sigma_1}
+12\,\coshx{2 \lambda_1}\\
&&+36\,\coshx{2 \lambda_1} \coshx{2 \lambda_2}
-\frac{9}{8}\,\coshx{ \lambda_1} \coshx{4 \lambda_2} \coshx{ \sigma_1}\\
&&-\frac{9}{8}\,\coshx{ \lambda_1} \coshx{ \sigma_1} \cosx{4 \varphi}
+\frac{93}{2}\,\coshx{ \lambda_1} \coshx{2 \lambda_2} \coshx{ \sigma_1}\\
&&+\frac{405}{8}\,\coshx{ \lambda_1} \coshx{ \sigma_1}
+\frac{7}{2}\,\coshx{4 \lambda_2}
+\frac{1}{2}\,\coshx{4 \lambda_2} \cosx{4 \varphi}\\
&&+14\,\coshx{2 \lambda_2}
-2\,\coshx{2 \lambda_2} \cosx{4 \varphi}
\end{array}
\end{equation}
where we again find independence of $\vartheta$ and $\Psi$.%
\footnote{Even if one uses $\vartheta,\Psi$-independence from start,
the head-on calculation in explicit component notation produces as an
intermediate quantity an (admittedly not maximally reduced) $T$-tensor
containing $83192$ summands which in turn contain $550148$
trigonometric functions, not counting powers; today, with some careful
programming, this is quite manageable for a desktop machine, but it
clearly shows the futility of this approach if one were to do such
calculations by hand.}  It must be emphasized that despite its
complexity this is still not the potential on the {\em complete}
subspace of $SU(3)$ singlets, since the $F$ and $S_2$ singlets have
not been included yet.

This potential defies a complete analysis on the symbolic level using
technology available today. Nevertheless, it is possible to extract
further candidates for stationary points by either employing numerics
or making educated guesses at the values of some coordinates. Here, we
take (as explained above) $(a,\phi)\in\left\{0,
\frac{1}{2}\pi,\pi,\frac{3}{2}\pi\right\}\times
\left\{0, \frac{1}{4}\pi, \frac{1}{2}\pi, \frac{3}{4}\pi\right\}$ and thus obtain the following six cases:
\vfill

\begin{enumerate}


\item $a=\pi, \phi=\frac{1}{2}\pi$:

\begin{equation}
\begin{array}{lll}
-8g^{-2}V&=&84 -\coshx{3 \lambda_1+ \sigma_1} \coshx{4 \lambda_2}
+\coshx{3 \lambda_1+ \sigma_1}\\
&&+24\,\coshx{2 \lambda_1}
+24\,\coshx{2 \lambda_1} \coshx{2 \lambda_2}\\
&&-3\,\coshx{ \lambda_1-\sigma_1} \coshx{4 \lambda_2}
+48\,\coshx{ \lambda_1} \coshx{2 \lambda_2} \coshx{ \sigma_1}\\
&&+24\,\coshx{ \lambda_1+ \sigma_1}
+27\,\coshx{ \lambda_1- \sigma_1}\\
&&+4\,\coshx{4 \lambda_2}
+24\,\coshx{2 \lambda_2}
\end{array}
\end{equation}


\item $a=\pi, \phi\in\left\{\frac{1}{4}\pi, \frac{3}{4}\pi\right\}$:

\begin{equation}
\begin{array}{lll}
-8g^{-2}V&=&93
-\frac{9}{4}\,\coshx{3 \lambda_1+ \sigma_1} \coshx{4 \lambda_2}
-\coshx{3 \lambda_1 +  \sigma_1} \coshx{2 \lambda_2}\\
&&+\frac{13}{4}\,\coshx{3 \lambda_1+ \sigma_1}
+12\,\coshx{2 \lambda_1}
+36\,\coshx{2 \lambda_1} \coshx{2 \lambda_2}\\
&&-\frac{3}{4}\,\coshx{ \lambda_1- \sigma_1} \coshx{4 \lambda_2}
+36\,\coshx{ \lambda_1+ \sigma_1} \coshx{2 \lambda_2}\\
&&+9\,\coshx{ \lambda_1- \sigma_1} \coshx{2 \lambda_2}
+12\,\coshx{ \lambda_1 + \sigma_1}\\
&&+\frac{159}{4}\,\coshx{ \lambda_1- \sigma_1}
+3\,\coshx{4 \lambda_2}
+16\,\coshx{2 \lambda_2}
\end{array}
\end{equation}


\item $a=\pi, \phi=0$:

\begin{equation}
\begin{array}{lll}
-8g^{-2}V&=&108
-4\,\coshx{3 \lambda_1+ \sigma_1} \coshx{4 \lambda_2}
+4\,\coshx{3 \lambda_1+ \sigma_1}\\
&&+48\,\coshx{2 \lambda_1} \coshx{2 \lambda_2}
+48\,\coshx{ \lambda_1+ \sigma_1} \coshx{2 \lambda_2}\\
&&+48\,\coshx{ \lambda_1- \sigma_1}
+4\,\coshx{4 \lambda_2}
\end{array}
\end{equation}


\item $a\in\left\{\frac{1}{2}\pi, \frac{3}{2}\pi\right\}, \phi=\frac{1}{2}\pi$:

\begin{equation}
\begin{array}{lll}
-8g^{-2}V&=&108
-4\,\coshx{3 \lambda_1} \coshx{4 \lambda_2} \coshx{ \sigma_1}
+4\,\coshx{3 \lambda_1} \coshx{ \sigma_1}\\
&&+48\,\coshx{2 \lambda_1} \coshx{2 \lambda_2}
+48\,\coshx{ \lambda_1} \coshx{2 \lambda_2} \coshx{ \sigma_1}\\
&&+48\,\coshx{ \lambda_1} \coshx{ \sigma_1}
+4\,\coshx{4 \lambda_2}
\end{array}
\end{equation}

\vfill\break


\item $a\in\left\{\frac{1}{2}\pi, \frac{3}{2}\pi\right\}, \phi\in\left\{\frac{1}{4}\pi, \frac{3}{4}\pi\right\}$:

\begin{equation}
\begin{array}{lll}
-8g^{-2}V&=&93
-\frac{9}{4}\,\coshx{3 \lambda_1} \coshx{4 \lambda_2} \coshx{ \sigma_1}
-\coshx{3 \lambda_1} \coshx{2 \lambda_2} \coshx{ \sigma_1}\\
&&+\frac{13}{4}\,\coshx{3 \lambda_1} \coshx{ \sigma_1}
+12\,\coshx{2 \lambda_1}
+36\,\coshx{2 \lambda_1} \coshx{2 \lambda_2}\\
&&-\frac{3}{4}\,\coshx{ \lambda_1} \coshx{4 \lambda_2} \coshx{ \sigma_1}
+45\,\coshx{ \lambda_1} \coshx{2 \lambda_2} \coshx{ \sigma_1}\\
&&+\frac{207}{4}\,\coshx{ \lambda_1} \coshx{ \sigma_1}
+3\,\coshx{4 \lambda_2}
+16\,\coshx{2 \lambda_2}
\end{array}
\end{equation}


\item $a\in\left\{\frac{1}{2}\pi, \frac{3}{2}\pi\right\}, \phi=0$:
\begin{equation}
\begin{array}{lll}
-8g^{-2}V&=&84
-\coshx{3 \lambda_1} \coshx{4 \lambda_2} \coshx{ \sigma_1}
+\coshx{3 \lambda_1} \coshx{ \sigma_1}\\
&&+24\,\coshx{2 \lambda_1}
+24\,\coshx{2 \lambda_1} \coshx{2 \lambda_2}
-3\,\coshx{ \lambda_1} \coshx{4 \lambda_2} \coshx{ \sigma_1}\\
&&+48\,\coshx{ \lambda_1} \coshx{2 \lambda_2} \coshx{ \sigma_1}
+51\,\coshx{ \lambda_1} \coshx{ \sigma_1}\\
&&+4\,\coshx{4 \lambda_2}
+24\,\coshx{2 \lambda_2}
\end{array}
\end{equation}
\end{enumerate}

The case $a=0,\pi=\frac{1}{2}\pi$ gives just the same potential as
$a=\pi, \phi=\frac{1}{2}\pi$, but with
$\sigma_1\leftrightarrow-\sigma_1$.  Furthermore,
$a=0,\phi\in\left\{\frac{1}{4}\pi, \frac{3}{4}\pi\right\}$ and $a=0,
\phi=0$ correspond to $a=\pi,\phi\in\left\{\frac{1}{4}\pi,
\frac{3}{4}\pi\right\}$, respectively $a=\pi,\phi=0$, both with 
$\sigma_1\leftrightarrow-\sigma_1$. In each of the cases $(2), (3),
(6)$, a detailed analysis produces a subcase of unmanageable
complexity; aside from these, cases $(1), (3), (4)$ feature nontrivial
stationary points that turn out to be true solutions of
eq. (\ref{SolutionSieve}).

\section{Five Extrema}

Many of the extrema of the potentials listed in the last section
re-appear multiple times; of every set of coordinates connected by various
sign flips or coordinate degeneracies, we only list one representative.

\begin{landscape}
\begin{figure}
\[
\begin{array}{|c|c|c|c|c|c|}
\hline
\mbox{Extremum}&\mbox{Location}&\mbox{Form}&\mbox{Cosmological}&\mbox{Remaining}&\mbox{Remaining}\\
\mbox{}&(\sigma_1,\lambda_1,\lambda_2,a,\phi)&\mbox{of the}&\mbox{constant}&\mbox{group}&\mbox{super-}\\
\mbox{}&\mbox{}&\mbox{scalar field}&\mbox{$\Lambda=4V$}&\mbox{symmetry}&\mbox{symmetry}\\
\hline
X_1 & (-K_1,K_1,K_1,0,0) & \exp\left(K_1\,\left(G_1^++G_2^+-S_1\right)\right) & -200\,g^2 & SO(7)^+ \times SO(7)^+ & \rm{None}\\
&&&&&\\
X_2 & (-K_2,K_2,-2\,K_1,\pi,\frac{\pi}{2}) & \exp\left(-K_2\,\left(G_1^++S_1\right)-2K_1\,G_2^-\right) & -416\,g^2 & SU(4) & \rm{None}\\
&&&&&\\
X_3 & (-K_3, -K_3, -K_3, \pi, \frac{\pi}{2}) &  \exp\left(K_3\,\left(G_1^+-S_1- G_2^-\right)\right) & -288\,g^2 & SU(3)\times SU(3)\times & (n_L,n_R)=(2,2)\\
&&&&\times U(1)\times U(1) &\\
&&&&&\\
X_4 & (K_6,K_7,K_9,\pi, \frac{\pi}{2}) & \exp\left(-K_7\,G_1^+ + K_9\,G_2^- + K_6\,S_1\right) & K_{10}\,g^2 & SU(3)\times U(1)\times U(1) & \rm{None}\\
&&&&&\\
X_5 & (-2\,K_2,0,-2\,K_1,\frac{\pi}{2},\frac{\pi}{2})& \exp\left(-2K_2\,S_1 -2K_1\,G_2^-\right) & -416\,g^2 & SU(4)^- & \rm{None}\\
\hline
\end{array}
\]
where
\[
\begin{array}{cclcccl}
K_1&=&\frac{1}{4}\,\ln\left(7+4\,\sqrt{3}\right)\approx0.6584789 &\qquad&
K_2&=&\frac{1}{2}\,\ln\left(\frac{5}{2}+\frac{1}{2}\,\sqrt{21}\right)\approx0.7833996\\
K_3&=&\ln(1+\sqrt{2})\approx0.8813736&&
K_4&=&\sqrt{6+6\,\sqrt{33}}\approx6.3613973\\
K_5&=&18+6\,\sqrt{33}+6\,K_4\approx90.6357598&&
K_6&=&\ln\Bigl(\frac{1}{18}\sqrt{K_5}\left(19+\frac{5}{3}\sqrt{33}+\frac{5}{3}\,K_4+\frac{1}{144}\,K_5^2-\frac{1}{7776}\,K_5^3\right)\Bigr)\\&&&&&&\approx-1.3849948\\
K_7&=&\ln\left(\frac{1}{6}\,\sqrt{K_5}\right)\approx0.4616649&&
K_8&=&\sqrt{78+14\sqrt{33}}\approx12.5866547\\
K_9&=&\frac{1}{2}\,\ln\left(\frac{7}{2}+\frac{1}{2}\,\sqrt{33}-\frac{1}{2}\,K_8\right)\approx-1.2694452\\
K_{10}&=&\multicolumn{5}{l}{-28512\left(1453+253\sqrt{33}+116\,K_8+20\sqrt{33}\,K_8\right)\,\left(3+\sqrt{33}+K_4\right)^{-2}\left(7+\sqrt{33}+K_8\right)^{-2}}\\
&&\left(24+6\sqrt{33}-3\,K_4-\sqrt{33}\,K_4\right)^{-1}\approx-398.5705673
\end{array}
\]
\caption{The Extrema}
\end{figure}
\end{landscape}

The extremum $X_1$ breaks the diagonal $SO(8)$ of $SO(8)\times SO(8)$
down to a $SO(7)$ under which the $SO(8)$ spinor decomposes as ${\bf
8}\rightarrow{\bf 7}+{\bf 1}$. In the notation of \cite{Warner:vz}
this is called $SO(7)^+$; it is easily checked that both $SO(8)$ of
$SO(8)\times SO(8)$ are broken in the same way. Hence this extremum
corresponds to $E_2$ in \cite{Warner:vz}. The extrema $X_1, X_2, X_3$
all break the diagonal $SO(8)$ down to $SU(3)\times U(1)$ and hence
loosely correspond to $E_5$ of \cite{Warner:vz}. It is easy to check
that the remaining symmetry of $X_4$ is 10-dimensional and hence has
to be $SU(3)\times U(1)\times U(1)$. Likewise, the remaining symmetry
of $X_3$ is 18-dimensional, and since its derivative is
16-dimensional, it has to be $SU(3)\times SU(3)\times U(1)\times
U(1)$. $X_5$ breaks the diagonal $SO(8)$ down to $SU(4)^-$ (again
using the nomenclature of \cite{Warner:vz}), and since the remaining
symmetry is only 15-dimensional, this is all that remains. This
extremum corresponds to $E_4$ of \cite{Warner:vz}. There is strong
evidence that the extrema $X_2$ and $X_5$ indeed are equivalent,
despite breaking the {\em diagonal} $SO(8)$ to different subgroups; a
detailed examination of these stationary points will have to show
whether this is really the case and what this might tell us about the
$D=4$ vacua $E_4$ and $E_5$. Furthermore, it is a bit unexpected to
see all cosmological constants except one have rational values.

\section{Acknowledgments}

It is a pleasure to thank my supervisor H.~Nicolai for introducing me
to this problem and for helpful and encouraging comments as well as
H.~Samtleben for helpful discussions.

Furthermore, I want to thank M.~Lindner as well as the administrative
staff of the CIP computer pool of the physics department at the
Universit\"at M\"unchen for allowing me to do some of the calculations
on their machines.

\newpage
\appendix
\renewcommand{\theequation}{\Alph{section}.\arabic{equation}}
\renewcommand{\thesection}{Appendix \Alph{section}:}

\section{$E_{8(+8)}$ and $E_{7(+7)}$ conventions}

For quick reference, we assemble all used conventions in this
appendix.

Structure constants of $E_{8(+8)}$ are explicitly given as follows:
using the conventions of \cite{Green:sp}, we define
\begin{equation}
\begin{array}{ll}
\sigma_1=\left(\begin{array}{rr}1&0\\0&1\end{array}\right)&
\sigma_x=\left(\begin{array}{rr}0&1\\1&0\end{array}\right)\\
\sigma_z=\left(\begin{array}{rr}1&0\\0&-1\end{array}\right)&
\sigma_e=\left(\begin{array}{rr}0&1\\-1&0\end{array}\right)
\end{array}
\end{equation}
to construct $SO(8)$ $\Gamma$-matrices $\Gamma^a_{\alpha\dot\beta}$ via
\begin{equation}
\begin{array}{ll}
\Gamma^1=\sigma_e\times\sigma_e\times\sigma_e&
\Gamma^2=\sigma_1\times\sigma_z\times\sigma_e\\
\Gamma^3=\sigma_e\times\sigma_1\times\sigma_z&
\Gamma^4=\sigma_z\times\sigma_e\times\sigma_1\\
\Gamma^5=\sigma_1\times\sigma_x\times\sigma_e&
\Gamma^6=\sigma_e\times\sigma_1\times\sigma_x\\
\Gamma^7=\sigma_x\times\sigma_e\times\sigma_1&
\Gamma^8=\sigma_1\times\sigma_1\times\sigma_1.
\end{array}
\end{equation}

With the decomposition $I=(\alpha,\dot\beta)$ for $SO(16)$
vector indices in terms of $SO(8)$ indices as well as 
$A=(\alpha\dot\beta,ab)$ for spinor and
$\dot A=(\alpha a,b\dot\beta)$ for co-spinor indices,
we define $SO(16)$ $\Gamma$-matrices $\Gamma^I_{A\dot A}$ following
the conventions of \cite{Nicolai:1986jk}:
\begin{equation}
\begin{array}{ll}
\Gamma^\alpha_{\beta\dot\gamma\;\delta b}=\delta_{\beta\delta}\Gamma^b_{\alpha\dot\gamma}&
\Gamma^\alpha_{ab\;c\dot\delta}=\delta_{ac}\Gamma^b_{a\dot\delta}\\
\Gamma^{\dot\alpha}_{ab\;\beta c}=\delta_{bc}\Gamma^a_{\beta\dot\alpha}&
\Gamma^{\dot\alpha}_{\beta\dot\gamma\;b\dot\delta}=-\delta_{\dot\gamma\dot\delta}\Gamma^b_{\beta\dot\alpha}.
\end{array}
\end{equation}

Splitting $E_8$ indices $\mathcal{A}, \mathcal{B},\ldots$ via
$\mathcal{A}=(A,[IJ])$, the structure constants of $E_{8(+8)}$ are
given by
\begin{equation}
f_{IJ\;KL}{}^{MN}=-8\,\delta_{{}_[I{}^[K}^{\phantom{MN}}\delta_{L{}^]J{}_]}^{MN},\qquad
f_{IJ\,A}{}^{B}={\textstyle \frac{1}{2}}\,\Gamma^{IJ}_{AB}.
\end{equation}
following the conventions of \cite{Nicolai:2000sc, Nicolai:2001sv}.

The common convention that for every antisymmetric index pair $[IJ]$
that is summed over, an extra factor $1/2$ has to be
introduced more explicitly corresponds to splitting $E_{8(+8)}$
indices not like $\mathcal{A}\rightarrow(A,[IJ])$, but instead like
$\mathcal{A}\rightarrow(A,\underline{[IJ]})\rightarrow(A,[IJ])$, where
$\underline{[IJ]}$ is treated as a single index in the range
$1\ldots120$ (and hence only summed over once) and the split
$\underline{[IJ]}\rightarrow[IJ]$ is performed using the map
$M^{\underline{[IJ]}}{}_{KL}=2\,\delta^{IJ}_{KL}$. If we sum over
$[IJ]$ after this split, we have to include a factor $1/2$.

From the generators
$t_\mathcal{A}{}^{\mathcal{C}}{}_{\mathcal{B}}=f_{\mathcal{A}\mathcal{B}}{}^{\mathcal{C}}$,
we form the Cartan-Killing metric
\begin{equation}
\eta_{\mathcal{A}\mathcal{B}}={\textstyle \frac{1}{60}}\tr t_\mathcal{A}t_\mathcal{B};\qquad\eta_{AB}=\delta_{AB},\quad\eta_{IJ\;KL}=-2\delta^{IJ}_{KL}.
\end{equation}

If we further use lexicographical order for $[IJ]$ index pairs,
taking only $I<J$ and use as the canonical $SO(16)$ Cartan subalgebra
\[t_{[1\,2]}=T_{129}, t_{[3\,4]}=T_{158},\ldots,t_{[15\,16]}=T_{248},\]
then the generators corresponding to the simple roots of $E_8$ are explicitly 
\begin{equation}
\begin{array}{lll}
T_{\scriptscriptstyle{+------+}}&=&T_{35}+i\,T_{36}+i\,T_{43}-T_{44}\\
T_{\scriptscriptstyle{+2-3}}&=&T_{159}-i\,T_{160}+i\,T_{171}+T_{172}\\
T_{\scriptscriptstyle{+3-4}}&=&T_{184}-i\,T_{185}+i\,T_{194}+T_{195}\\
T_{\scriptscriptstyle{+4-5}}&=&T_{205}-i\,T_{206}+i\,T_{213}+T_{214}\\
T_{\scriptscriptstyle{+5-6}}&=&T_{222}-i\,T_{223}+i\,T_{228}+T_{229}\\
T_{\scriptscriptstyle{+6-7}}&=&T_{235}-i\,T_{236}+i\,T_{239}+T_{240}\\
T_{\scriptscriptstyle{+7-8}}&=&T_{244}-i\,T_{245}+i\,T_{246}+T_{247}\\
T_{\scriptscriptstyle{+7+8}}&=&T_{244}+i\,T_{245}+i\,T_{246}-T_{247}.
\end{array}
\end{equation}

The fundamental 56-dimensional matrix representation of the
$E_{7(+7)}$ Lie algebra decomposes into $28\times28$ submatrices
under its maximal compact subgroup $SU(8)$
\begin{equation}
\left(\begin{array}{cc}
2A_{[{\tt i}}{}^{[{\tt I}}\delta_{{\tt j}]}{}^{{\tt J}]}&\Sigma_{\tt ijKL}\\
\Sigma^{\tt klIJ}&2A_{[{\tt k}}{}^{[{\tt K}}\delta_{{\tt l}]}{}^{{\tt L}]}
\end{array}\right)
\end{equation}
where $A_{\tt i}{}^{\tt I}$ is an anti-hermitian traceless complex
$8\times8$ matrix generator of SU(8) and
$\Sigma_{\tt ijKL}=\overline{\Sigma^{\tt ijKL}}$
is complex, self-dual and totally antisymmetric.

We obtain this subalgebra from the $E_{8(+8)}$ algebra as follows: we
form $U(8)$ indices from $SO(16)$ indices via 
\begin{equation}
x^{\tt j}+i\,x^{({\tt j}+8)}=z^{\tt j},\qquad {\tt j}=1\ldots8
\end{equation}
and thus identify the corresponding $SU(8)$ subalgebra within
$SO(16)$. Under this embedding, $SU(8)$ generators are lifted to
$SO(16)$ via
\begin{equation}
\begin{array}{lll}
G^{\{SO(16)\}\tt i}{}_{\tt j}&=&\phantom+\Re\left(G^{\{SU(8)\}}{}_{\tt i}{}^{\tt j}\right)\\
G^{\{SO(16)\}\tt i+8}{}_{\tt j+8}&=&\phantom+\Re\left(G^{\{SU(8)\}}{}_{\tt i}{}^{\tt j}\right)\\
G^{\{SO(16)\}\tt i+8}{}_{\tt j}&=&\phantom+\Im\left(G^{\{SU(8)\}}{}_{\tt i}{}^{\tt j}\right)\\
G^{\{SO(16)\}\tt i}{}_{\tt j+8}&=&-\Im\left(G^{\{SU(8)\}}{}_{\tt i}{}^{\tt j}\right)
\end{array}
\end{equation}
and then to $E_{8(+8)}$ by
\begin{equation}
t^{\mathcal M}{}_{\mathcal N} = f_{\underline{[IJ]}\mathcal{N}}{}^{\mathcal{M}} M^{\underline{[IJ]}}{}_{I}{}^{J}G^{\{SO(16)\}I}{}_{J}.
\end{equation}

Furthermore, we form raising and lowering operators from
$SO(16)$ $\Gamma$-matrices:
\begin{equation}
\Gamma^{{\tt j}\pm}=\frac{1}{2}\left(\Gamma^{\tt j}\pm i\,\Gamma^{{\tt j}+8}\right),\qquad {\tt j}=1\ldots8
\end{equation}

We call the $SO(16)$ Weyl-Spinor that is annihilated by all
$\Gamma^{{\tt j}-}$ $\psi_{\scriptscriptstyle 0}$. With our conventions, the complex
$E_8$ generator corresponding to this spinor is
\begin{equation}
\begin{array}{lcl}
T_{\scriptscriptstyle}&=&\frac{1}{4}\left(T_{1}+T_{10}+T_{19}+T_{28}+T_{37}+T_{46}+T_{55}+T_{64}\right.\\
&&\left.+i\left(T_{65}+T_{74}+T_{83}+T_{92}+T_{101}+T_{110}+T_{119}-T_{128}\right)\right).
\end{array}
\end{equation}

Since the charge-conjugation matrix is just the identity in our basis,
the real as well as the imaginary part of every spinor obtained by
acting with an even number of $\Gamma^{{\tt j}+}$ on
$\psi_{\scriptscriptstyle 0}$ is a Majorana-Weyl-spinor.
We use the terminology
\begin{equation}
\begin{array}{l}
\psip{{\tt i}_1 {\tt i}_2\ldots {\tt i}_{2n}}=\phantom+\mbox{Re}\,\Gamma^{{\tt i}_1+}\Gamma^{{\tt i}_2+}\cdots\Gamma^{{\tt i}_{2n}+}\psi_{\scriptscriptstyle 0}\\
\psim{{\tt i}_1 {\tt i}_2\ldots {\tt i}_{2n}}=-\mbox{Im}\,\Gamma^{{\tt i}_1+}\Gamma^{{\tt i}_2+}\cdots\Gamma^{{\tt i}_{2n}+}\psi_{\scriptscriptstyle 0}.
\end{array}
\end{equation}

Note that
\begin{equation}
\begin{array}{l}
\psip{}=\frac{1}{2}\,\left(\psi_{\scriptscriptstyle 0}-\psi_{1\ldots8}\right)\\
\psim{}=\frac{i}{2}\,\left(\psi_{\scriptscriptstyle 0}+\psi_{1\ldots8}\right)\\
\psip{{\tt i}_1 {\tt i}_2}=-\frac{1}{2}\,\left(\psi_{{\tt i}_1 {\tt i}_2}+\delta^{{\tt i}_1\ldots {\tt i}_8}_{1\ldots8}\,\psi_{{\tt i}_3\ldots {\tt i}_8}\right)\\
\psim{{\tt i}_1 {\tt i}_2}=-\frac{i}{2}\,\left(\psi_{{\tt i}_1 {\tt i}_2}-\delta^{{\tt i}_1\ldots {\tt i}_8}_{1\ldots8}\psi_{{\tt i}_3\ldots {\tt i}_8}\right)\\
\psip{{\tt i}_1 {\tt i}_2 {\tt i}_3 {\tt i}_4}=\frac{1}{2}\,\left(\psi_{{\tt i}_1\ldots {\tt i}_4}-\delta^{{\tt i}_1\ldots {\tt i}_8}_{1\ldots8}\,\psi_{{\tt i}_5\ldots {\tt i}_8}\right)\\
\psim{{\tt i}_1 {\tt i}_2 {\tt i}_3 {\tt i}_4}=\frac{i}{2}\,\left(\psi_{{\tt i}_1\ldots {\tt i}_4}+\delta^{{\tt i}_1\ldots {\tt i}_8}_{1\ldots8}\,\psi_{{\tt i}_5\ldots {\tt i}_8}\right).
\end{array}
\end{equation}

This way, the 35 $E_{8(+8)}$ generators
$t_{A}{}^{\mathcal{C}}{}_{\mathcal{B}}\psi^A$ corresponding to the
$\psim{{\tt i}_1 {\tt i}_2 {\tt i}_3 {\tt i}_4}$ are the $E_{7(+7)}$ generators for
the real parts of the $\Sigma_{{\tt i}_1 {\tt i}_2 {\tt i}_3 {\tt i}_4}$ while the generators
corresponding to $\psip{{\tt i}_1 {\tt i}_2 {\tt i}_3 {\tt i}_4}$ carry the
imaginary parts.

All in all, under the $E_{7(+7)}\times SL(2)\subset E_{8(+8)}$
embedding considered here the ${\bf 248}=({\bf 133},{\bf 1})+({\bf
56},{\bf 2})+({\bf 1},{\bf 3})$ generators of $E_{8(+8)}$ which
decompose into $120$ compact generators corresponding to the adjoint
representation of $SO(16)$ and $128$ noncompact Majorana-Weyl-Spinors
of $SO(16)$ further decompose as follows: the compact $63$ $SU(8)$
generators form the maximal compact subgroup of the $({\bf 133},{\bf 1})$ adjoint
representation of $E_{7(+7)}$ while the extra $U(1)$ generator
provides the compact generator of the adjoint $SL(2)$ representation
$({\bf 1},{\bf 3})$. The $2\times 35$ real spinors $\psi^{\pm}_{{\tt i}_1 {\tt
i}_2 {\tt i}_3 {\tt i}_4}$ provide the $70$ noncompact generators in
$({\bf 133},{\bf 1})$ while the spinors $\psi^{\pm}_{}$ give the two noncompact
generators in $({\bf 1},{\bf 3})$. The remaining $56$ compact generators group
together with the $\psipm{{\tt i}_1 {\tt i}_2}$ spinors and form the
$({\bf 56},{\bf 2})$ fundamental representation of $E_{7(+7)}\times SL(2)$.

\newpage

\begingroup\raggedright\endgroup

\end{document}